\documentclass[prb,aps,epsf,twocolumn,superscriptaddress,showpacs,10pt]{revtex4-2}
\usepackage{graphicx,amsfonts,times,bm,amsmath,verbatim,color,array}
\usepackage{lineno}
\usepackage{xcolor}
\usepackage{algorithmic}
\usepackage{float}
\usepackage[colorlinks,urlcolor=blue,citecolor=blue,linkcolor=blue]{hyperref}
\newcommand{\bra}[1]{\langle {#1} |}
\newcommand{\ket}[1]{| {#1} \rangle}

\begin{document}

\title{Nonlinear anomalous transverse responses induced by Berry curvature quadrupole in systems with broken time-reversal symmetry}
\author{Srimayi Korrapati}
\affiliation{Department of Physics and Astronomy, Clemson University, Clemson, SC 29634, USA}

\author{Snehasish Nandy}
\affiliation{Department of Physics, National Institute of Technology Silchar, Assam, 788010, India}
\affiliation{Theoretical Division, Los Alamos National Laboratory, Los Alamos, New Mexico 87545, USA}

\author{Sumanta Tewari}
\affiliation{Department of Physics and Astronomy, Clemson University, Clemson, SC 29634, USA}

\begin{abstract}
Recent theoretical work has shown that higher-order moments of the Berry curvature, e.g., Berry curvature quadrupole and hexapole moments, can produce the leading order nonlinear anomalous Hall response (NLAH) in systems with special magnetic point group symmetry. Recent experimental work has reported the observation of the Berry curvature quadrupole-induced third-order NLAH (i.e., Hall voltage proportional to the third power of the external electric field) from cryogenic conditions to room temperature in an epitaxially grown material platform with broken time-reversal symmetry. In this paper, using semiclassical Boltzmann formalism in the relaxation time approximation, we compute the Berry curvature quadrupole-induced nonlinear anomalous thermal Hall and Nernst coefficients in time-reversal broken systems. In systems where Berry curvature monopole and dipole moments vanish by symmetry, our results predict the behavior of the leading order anomalous thermal Hall and Nernst coefficients proportional to the third power of the applied longitudinal temperature gradient. They are guaranteed to exist in systems that have already exhibited the third-order nonlinear anomalous Hall effect in recent experiments.        
\end{abstract}

\maketitle

\section{\label{seclevel1}INTRODUCTION}
Topological transport phenomena via observation of transverse charge, spin, or thermoelectric effects in response to an applied longitudinal electric field or temperature gradient have recently attracted immense attention.  In the conventional electric charge Hall effect, a transverse electric voltage is generated in response to an applied electric field/current in the presence of an out-of-plane magnetic field providing Lorentz force and deflecting the charge carriers~\cite{Hall_1879}. With non-trivial geometric properties of the Bloch wave functions such as Berry curvature~\cite{Xiao_2010}, a transverse electric voltage proportional to the first power of the applied electric field can be non-zero even in the absence of an applied out-of-plane magnetic field~\cite{Karplus_1954, Luttinger_1958, Jungwirth_2002, Nagaosa_2010}. This effect called the anomalous charge Hall effect, and \textcolor{black}{the associated counterparts induced by a temperature gradient, such as anomalous thermal Hall and anomalous Nernst effects}, are non-zero in systems where the integral of the Berry curvature over the first Brillouin zone, called Berry curvature monopole (BCM), is non-zero over the occupied bands~\cite{Niu_2006, Xiao_2010, Bergman_2010, Nagaosa_2010, Roy_2022}. The BCM-induced topological charge, spin, and thermoelectric transport phenomena have been a fascinating field of study over the past two decades and have been experimentally observed in systems with non-trivial band structures~\cite{Taguchi_2001, Fang_2003, Ong_2004, Lee_2004, Ikhlas_2017, Li_2017, Wuttke_2019, Guin_2019, Sakai_2020}.  

In first-order charge current response to an applied electric field ($j_{\alpha}=\sigma_{\alpha \beta}E_{\beta}$, where $\alpha,\beta =x,y$ indicate spatial directions), the antisymmetric part of the conductivity tensor  ($\sigma^{H}_{\alpha \beta}=-\sigma^{H}_{\beta \alpha}$) is the Hall conductivity and is non-dissipative. The existence of the first-order non-dissipative Hall conductivity requires systems with broken time-reversal symmetry (TRS)~\cite{Onsagar_1931, Nagaosa_2010}. 
In the nonlinear current regime ($j_{\alpha}=\chi_{\alpha\beta\gamma}E_{\beta}E_{\gamma}$), on the other hand, a non-dissipative Hall voltage proportional to the second power of the applied electric field can be non-zero even in time-reversal-symmetric systems~\cite{Sodemann_2015}. In time-reversal symmetric systems with broken space inversion symmetry, the dissipationless second-order Hall current can result from a non-zero first-order moment of the Berry curvature, the so-called Berry curvature dipole (BCD), defined as the integral of the first-order derivative of the Berry curvature over the first Brillouin zone of the occupied bands~\cite{Sodemann_2015}. The Berry curvature dipole-induced second-order anomalous Hall effect has recently been observed in experiments~\cite{Ma_2019, Mak_2019, Du_2021}. Additionally, the analogous anomalous thermoelectric effects, e.g., BCD-induced nonlinear anomalous thermal Hall and Nernst effects have also been theoretically predicted for systems supporting BCD-induced second-order nonlinear Hall effect~\cite{Gao_2018, Yu_2019, Nagaosa_2019, Zeng_2019, Zeng_2020, Zeng_2021}.

Both the Berry curvature monopole and dipole moments are subject to symmetry constraints, violations of which force them to vanish~\cite{Xiao_2010, Du_2021}. In particular, all the odd-order responses induced by Berry curvature multipoles (e.g., BCM-induced transverse voltage response proportional to the \textit{first power} of the applied electric field) require broken time-reversal symmetry, and the corresponding even-order responses (e.g., BCD-induced transverse voltage response proportional to the \textit{second power} of the applied electric field) require broken space inversion symmetry. Furthermore, there are additional crystallographic symmetry constraints based on which it has been predicted recently that in fifteen three-dimensional (3D) magnetic space groups and three two-dimensional (2D) magnetic space groups, both Berry curvature monopole and dipole moments identically vanish~\cite{Zhang_2023}. In these systems, the leading-order anomalous Hall voltage in response to an applied electric field $E_x$ is proportional to the \textit{third power} of $E_x$ and is governed by the Berry curvature quadrupole moment on the Fermi surface defined as the integral of the second-order derivative of the Berry curvature over the first Brillouin zone of the occupied bands~\cite{fang2023quantum}. Remarkably, the experimental observation of such a third-order anomalous Hall voltage in response to an applied longitudinal current has also been reported recently, pointing strongly to a topological origin of this effect originating from a nonzero Berry curvature quadrupole moment~\cite{Sankar_2023}. 

In this paper, using semiclassical Boltzmann transport formalism in the relaxation time approximation, augmented by Berry curvature-induced anomalous terms in the semiclassical equations of motion, we analytically derive the general expression of the energy-modulated Berry curvature quadrupole-induced nonlinear anomalous thermal Hall and Nernst coefficients in time-reversal broken systems. We find that both thermal Hall and Nernst coefficients scale quadratically with the relaxation time. Considering a two-dimensional (2D) Rashba-like system with second-order Fermi surface warping, we show the energy-modulated Berry curvature quadrupole can appear as a leading-order response. We also show the variation of third-order anomalous thermal Hall and Nernst effect in this system as a function of chemical potential and temperature. It is important to note that in systems where Berry curvature monopole and dipole moments vanish by symmetry, e.g., the fifteen 3D and three 2D magnetic space group systems~\cite{Zhang_2023}, our results predict the behavior of the leading order transverse anomalous thermal Hall and Nernst coefficients proportional to the third power of the applied longitudinal temperature gradient. In addition, the quantities we calculate are guaranteed to be nonzero in systems that have already exhibited the third-order nonlinear anomalous Hall effect in recent experiments.

The remainder of the paper is organized as follows: in Sec.~\ref{formalism}, we derive the general expressions of third-order Nernst effect and thermal Hall effect within the framework of quasiclassical Boltzmann formalism and discuss symmetry considerations. In Sec.~\ref{results} we investigate the third-order anomalous transverse responses in response to a longitudinal thermal gradient in a 2D Rashba model system and discuss their dependencies on chemical potential and temperature. Finally, we conclude by summarizing the important results in Sec.~\ref{conclusions}.  

\section{\label{formalism}Quasiclassical Framework For third-order Thermoelectric coefficients}
In this section, we derive the general expression for third-order anomalous thermal Hall and Nernst coefficients in the diffusive regime within the semiclassical Boltzmann transport framework. The phenomenological Boltzmann transport equation can be written as~\cite{Ashcroft_1976, Ziman_2001}
\begin{equation}
\label{eqbte}
(\partial_{t}+\bm{\dot{r}}.{\bm{\nabla_r}}+\bm{\dot{k}}.{\bm{\nabla_k}})f(\bm{r},\bm{k},t) = I_{coll}\{f(\bm{r},\bm{k},t)\},
\end{equation}
where $f(\bm{r},\bm{k},t)$ denotes the local non-equilibrium electron distribution function, and $I_{coll}\{f(\bm{r},\bm{k},t)\}$ is the collision integral
which incorporates the effects of electron correlations and
impurity scattering. Within the relaxation time approximation, the steady-state Boltzmann transport equation takes the form
\begin{equation}\label{eqblt}    (\bm{\dot{r}}.\nabla_{\bm{r}}+\bm{\dot{k}}.\nabla_{\bm{k}})f(\bm{r},\bm{k}) = \frac{f_{0}(\bm{r},\bm{k})-f(\bm{r},\bm{k})}{\tau (\bm{k})},
\end{equation}
where $f_{0}(\bm{r},\bm{k}) = 1/(1+e^{\beta(\epsilon_{\bm{k}} - \mu)})$ is the local equilibrium distribution function in the absence of external fields with
an inhomogeneous temperature $\beta (\bm{r})=1/k_B T (\bm{r})$; $\epsilon_{\bm{k}}$ is the energy, $\mu$ denotes chemical potential. Here, $\tau (\bm{k})$ represents the average scattering time between two successive collisions. In this work, we ignore the momentum dependence of $\tau (\bm{k})$ to simplify the calculations and assume it to be a constant~\cite{PhysRevB.88.104412, PhysRevB.89.195137, PhysRevB.90.165115, Nandy_2023}.

For electrons moving in a periodic potential with momentum $\bm{k}$ and energy $\epsilon_{\bm{k}}$ corresponding to Bloch states $\ket{u_{\bm{k}}}$, the Berry curvature ${\bm{\Omega_k}}$ and Berry connection ${\bm{A_k}}$ are defined as \cite{Xiao_2010}
\begin{equation}
    \Omega_{\bm{k},a} \equiv \varepsilon_{abc}\partial_b A_c, \quad 
    A_{\bm{k},c} = -\bra{u_{\bm{k}}}\partial_c\ket{u_{\bm{k}}},
\end{equation}
where $a,b,c \in \{x,y,z\}$, $\varepsilon_{abc}$ is the Levi-Civita antisymmetric tensor, and $\partial_a = \partial /\partial k_a$. The Berry curvature associated with the Bloch states plays the role of a magnetic field in the momentum space and gives rise to interesting anomalous transport properties. Incorporating the Berry curvature effects, the semiclassical equations of electron motion in the absence of a magnetic field take the following form~\cite{Niu_1999, Niu_2006}
\begin{equation}
\label{eqv}
\bm{\dot{r}} = \frac{1}{\hbar}
    \frac{\partial\epsilon_{\bm{k}}}{\partial \bm{k}} - \frac{e}{\hbar}\bm{E \times \Omega_{\bm{k}}} , \quad \quad 
\bm{\dot{k}} = -\frac{e\bm{E}}{\hbar},
\end{equation}
where $v_a=\frac{1}{\hbar}
    \frac{\partial\epsilon_{\bm{k}}}{\partial k_a}$ is the group velocity. The second term in the expression for $\bm{\dot{r}}$ represents the anomalous velocity due to Berry curvature. 
In this work, we are interested in transverse anomalous coefficients in response to the third power of an applied temperature gradient ($-\bm{\nabla} T = -\partial T/\partial \bm{r}$) in the absence of an electric field. Substituting Eq.~(\ref{eqv}) with $\bm{E}=0$ into the Eq.~(\ref{eqblt}), the Boltzmann transport equation is simplified as
\begin{equation}
\label{eqbe2}
\bm{\dot{r}}.\bm{\nabla_r}f(\bm{r},\bm{k}) = \frac{f_{0}(\bm{r},\bm{k})-f(\bm{r},\bm{k})}{\tau},
\end{equation}
where we expand $f(\bm{r},\bm{k})$ up to the second-order as
\begin{equation}\label{eqf}
    f(\bm{r},\bm{k}) = \sum_{i=0}^{2} f_n (\bm{r},\bm{k}).
\end{equation}
Here, $f_n$ is understood as an $n$-th order response to the applied thermal gradient i.e., $f_n\propto(-\bm{\nabla} T)^n$. Plugging Eq.~(\ref{eqf}) in Eq.~(\ref{eqbe2}) and equating powers of $(-\bm{\nabla} T)$ gives rise to the recursive relation
\begin{equation}
    f_{n+1} = -\tau {\bm{\dot{r}}}.\bm{\nabla_r} f_n,
\end{equation}
which leads to expressions for the first and second-order terms of the distribution function
\begin{equation}\label{eqfE}
\begin{split}
    f_1 =& \frac{\tau}{\hbar}\nabla_a T \frac{(\epsilon_{\bm{k}}-\mu)}{T}\partial_a f_0,\\
    f_2 =& 2\frac{\tau^2}{\hbar} v_a \nabla_a T \nabla_b T \frac{(\epsilon_{\bm{k}} - \mu)}{T^2}\partial_b f_0\\
    &+\frac{\tau^2}{\hbar^2} \nabla_a T \nabla_b T \frac{(\epsilon_{\bm{k}} - \mu)^2}{T^2}\bigl(\partial^2_{ab} f_0 -\partial^2_{ab} \epsilon_{\bm{k}} \frac{\partial f_0}{\partial \epsilon_{\bm{k}}}\bigr)\\
    &-\frac{\tau^2}{\hbar} v_a \nabla^2_{ab} T \frac{(\epsilon_{\bm{k}} - \mu)}{T}\partial_b f_0,
 \end{split}
 \end{equation}
 where $\nabla_a = \partial/\partial r_a$. A repeated index implies the summation of the term over all the index values according to the Einstein notation. Here, we have used the following relations: 
 \begin{equation}
 \begin{split}
     \frac{\partial f_0}{\partial T}= & -\frac{(\epsilon_{\bm{k}} - \mu)}{T}\frac{\partial f_0}{\partial \epsilon_{\bm{k}}}, \quad 
     \frac{\partial^2 f_0}{\partial T^2}
     =\frac{\partial}{\partial \epsilon_{\bm{k}}}\biggl(\frac{(\epsilon_{\bm{k}} - \mu)^2}{T^2}\frac{\partial f_0}{\partial \epsilon_{\bm{k}}}\biggr),\\
     \partial_a f_0 = &\hbar v_a \frac{\partial f_0}{\partial \epsilon_{\bm{k}}}, \quad 
     \partial^2_{ab} f_0 = \hbar^2 v_a v_b \frac{\partial^2 f_0}{\partial \epsilon_{\bm{k}}^2}+ \partial^2_{ab} \epsilon_{\bm{k}} \frac{\partial f_0}{\partial \epsilon_{\bm{k}}}. 
 \end{split}
 \end{equation}

\subsection{\label{ATHE}Third-order Anomalous Thermal Hall Effect}
The anomalous thermal Hall effect refers to the Berry curvature-induced transverse thermal response to an applied longitudinal thermal gradient ($-\bm{\nabla} T$) in the absence of an external magnetic field~\cite{Niu_2006}. Considering all the conventional and anomalous contributions, the total thermal transport current density due to charge carriers is given by 
\begin{equation}
    \bm{j^Q} = \bm{j^Q_N}+\bm{j^Q_E}+ \bm{j^Q_T}, 
\end{equation}
where $\bm{j^Q_N}$ is the standard contribution coming from conventional group velocity $\bm{v_k}$ of the carriers, while $\bm{j^Q_E}$ and $\bm{j^Q_T}$ are mediated by the nontrivial Berry curvature in the presence of an electric field and thermal gradient, respectively. As a longitudinal response, $\bm{j^Q_N}$ does not contribute to the transverse responses we investigate in this work and is thus neglected. Additionally, in the current setup, $\bm{j^Q_E}$ vanishes due to the absence of an electric field.

In view of this, the total transverse thermal current density mediated by the nontrivial Berry curvature is given by \cite{Niu_2006, Bergman_2010, Murakami_2011} as  
\begin{equation}\label{eq:jqt}
\begin{split}
    \bm{j}^Q_T = & -\frac{k^2_{B}T}{\hbar}\bm{\nabla} T\times\int _{\bm{k}} \bm{\Omega}^m_{\bm{k}}\biggl[ \frac{(\epsilon^m_{\bm{k}}-\mu)^2}{(k_{B}T)^2} f^m_0 + \frac{\pi^2}{3} \\ &- \ln^2(1 + e^{-\beta(\epsilon^m_{\bm{k}}-\mu)}) -2\textrm{Li}_2(1-f^m_0) \biggr],
\end{split}
\end{equation}
where the integration is over the first Brillouin Zone, with  $\int_{\bm{k}} \equiv \sum_{m} \int d^{d}k/(2\pi)^{d}$ with $d$ being the dimensionality and $m$ being the band index; $\textrm{Li}_2(z)$ is the dilogarithm function. Substituting the solution for the non-equilibrium distribution function $f(\bm{r},\bm{k})$ in place of $f_0$ in Eq.~(\ref{eq:jqt}) allows us to write the $n$-th order ($n = 1, 2, 3$) thermal Hall current response in $a$-direction as
\begin{equation}\label{eq:jqtn}
    (j^Q_T)_{a,n} = -\varepsilon_{abc}\frac{k^2_{B}T}{\hbar}\nabla_bT \int_{\bm{k}}\Omega_{\bm{k},c}\frac{(\epsilon_{\bm{k}}-\mu)^2}{(k_{B}T)^2} f_{n-1},
\end{equation}
where we have dropped the band index for brevity. In this work, we have ignored the relatively small contributions arising from linearizing the dilogarithm term in Eq.~(\ref{eq:jqt}). Using the expression of $f_2$ given in Eq.~(\ref{eqfE}), 
the third-order anomalous thermal Hall current density is denoted as
\begin{equation}\label{eqj3}
\begin{split}
   (j^Q_T)_{a,3} =& - \kappa^{23}_{abde} (\nabla_b T) (\nabla^2_{de}T) \\
    &- \kappa^{222}_{abde} (\nabla_b T)(\nabla_d T)(\nabla_e T).
\end{split}
\end{equation}
The superscript indices of $\kappa$ specify the orders of the spatial derivatives of temperature $T$ that are involved in generating the heat current: $2$ for each external field that is a first derivative of the temperature ($\nabla_a T$), $3$ for each field that is a second derivative of the temperature ($\nabla^2_{ab} T$).

Now, the two conductivities associated with the anomalous third-order thermal Hall effect in Eq.(\ref{eqj3}) take the form
\begin{equation}\label{eq:k23}
    \kappa^{23}_{abde} = \varepsilon_{abc} \tau^2\frac{k_B^4 T^2}{\hbar^3} \lambda^{\kappa,2}_{cde},
    \end{equation}
and
\begin{equation}\label{eq:k222}
    \kappa^{222}_{abde} = -\varepsilon_{abc}\tau^2 \frac{k_B^4 T}{\hbar^3} (\lambda^{\kappa,1}_{cde} + 2\lambda^{\kappa,2}_{cde} - \lambda^{\kappa,3}_{cde}),
\end{equation}
\begin{figure}[htb]
\centering
\includegraphics[width=0.4\textwidth]{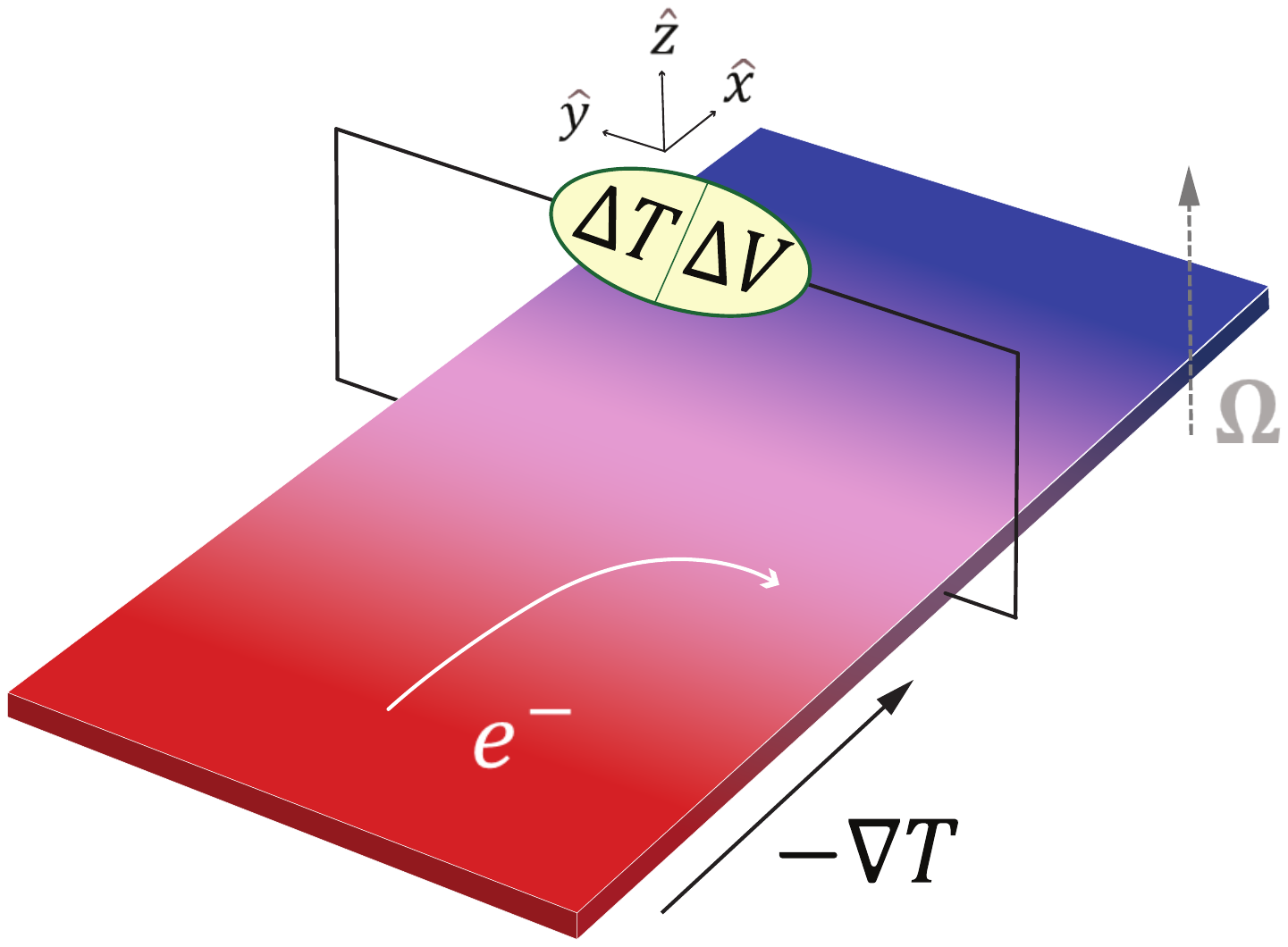}
\caption{Schematic diagram of the experimental setup for measuring nonlinear anomalous thermal Hall and Nernst effects. In the presence of a non-trivial Berry curvature ($\bm{\Omega}$), a longitudinal temperature gradient ($-\bm{\nabla} T$) gives rise to a transverse thermal gradient (thermal Hall effect $\Delta T$) and a transverse potential difference (Nernst effect $\Delta V$) even in the absence of an external magnetic field.}
\label{fig:exp}
\end{figure}
where
\begin{equation}\label{eq:lambda_kappa}
\begin{split}           
    \lambda^{\kappa,1}_{cde} = &\int_{\bm{k}} \Omega_{\bm{k},c} \frac{(\epsilon_{\bm{k}}-\mu)^4}{(k_{B}T)^4}  \bigl(\partial^2_{de} f_0)\\   
    \lambda^{\kappa,2}_{cde} = &\int_{\bm{k}}\Omega_{\bm{k},c}\frac{(\epsilon_{\bm{k}}-\mu)^3}{(k_{B}T)^4} (\partial_d \epsilon_{\bm{k}})(\partial_e f_0)\\
    \lambda^{\kappa,3}_{cde} = &\int_{\bm{k}}\Omega_{\bm{k},c}\frac{(\epsilon_{\bm{k}}-\mu)^4}{(k_{B}T)^4} \bigl(\partial^2_{de} \epsilon_{\bm{k}}) (\frac{\partial f_0}{\partial \epsilon_{\bm{k}}}\bigr).\\
\end{split}
\end{equation}
$\kappa^{23}_{abde}$ is generated by $\lambda^{\kappa,2}_{cde}$, and is one power of $T$ higher than $\kappa^{222}_{abde}$ which is seen to have contributions from all three $\lambda^{\kappa,n}_{cde}$'s. The expressions given in Eq.~(\ref{eq:lambda_kappa}) illustrate that all the $\lambda^{\kappa,n}$'s and, by extension, $\kappa^{23}$ and $\kappa^{222}$ are Fermi surface quantities due to the presence of $\partial f_0/\partial\epsilon_{\bm{k}}$ which reduces to $-\delta(\epsilon_{\bm{k}}-\mu)$ at zero temperature. Additionally, the conductivities scale quadratically with scattering time $\tau$.
In order to examine the low-temperature behavior of the two conductivities, we start by  rewriting $\lambda^{\kappa,2}_{cde}$ from Eq.~(\ref{eq:lambda_kappa}) as
\begin{equation}
    \lambda^{\kappa,2}_{cde} =\frac{1}{(k_BT)^4}\int d\epsilon f{'}(\epsilon) k_{cde}(\epsilon),
\end{equation}
where
\begin{equation}
    k_{cde}(\epsilon) = \int_{\bm{k}} \Omega_{\bm{k},c}\partial_{d}\epsilon_{\bm{k}}\partial_{e}\epsilon_{\bm{k}}(\epsilon-\mu)^3\delta(\epsilon-\epsilon_{k}).
\end{equation}
Following the Sommerfeld low-temperature expansion \cite{Ashcroft_1976}, we get
\begin{equation}
    \lambda_{cde}^{\kappa,2}(T,\mu) = \frac{1}{(k_BT)^4} \biggl[\frac{7\pi^4}{360}(k_BT)^4 k^{(4)}_{cde}(\mu) + O(T^6)+\dots \biggr],    
\end{equation}
where $k^{(n)}_{cde}(\mu) = \partial^{n}k_{cde}/\partial \mu^n$. Eq.~(\ref{eq:k23}) can now be rewritten as
\begin{equation}\label{eq:T_k23}
    \kappa^{23}_{abde}(T,\mu) = \varepsilon_{abc} \frac{7\pi^4}{360} \frac{k_B^4 T^2}{\hbar^3}\tau^2   k^{(4)}_{cde}(\mu) + O(T^4)+ \dots,
    \end{equation}
where we have $\kappa^{23}\propto T^2$ at low $T$. Similarly, we find that  $\lambda_{cde}^{\kappa,1}$ and $\lambda_{cde}^{\kappa,3}$ are proportional to $T^0$ in the leading order using the Sommerfeld expansion. Consequently, we obtain
\begin{equation}\label{eq:T_k222}
    \kappa^{222}_{abde}(T,\mu) = -\varepsilon_{abc} \frac{7\pi^4}{360}\frac{k_B^4 T}{\hbar^3}\tau^2  K^{(4)}_{cde}(\mu) + O(T^3) + \dots,  
\end{equation}
where 
\begin{equation}
    K_{cde}(\epsilon) = \int_{\bm{k}} \Omega_{\bm{k},c}\partial_{d}\epsilon_{\bm{k}}\partial_{e}\epsilon_{\bm{k}}\biggl[(\epsilon-\mu)^4\frac{\partial\delta}{\partial \epsilon} + 2(\epsilon-\mu)^3\delta(\epsilon-\epsilon_{k})\biggr],
\end{equation} 
with $K^{(n)}_{cde}(\mu) = \partial^{n}K_{cde}/\partial \mu^n$, and  $\kappa^{222}$ varies linearly with $T$ at low temperature.

To understand the origin of the anomalous third-order thermal Hall current, we examine the quadrupole moment of the Berry curvature over the occupied states 
\begin{equation}\label{eq:qorig}
        Q_{abc} = \int_{\bm{k}} (\partial^2_{ab}\Omega_{\bm{k},c}) f_0. 
\end{equation}
The Berry curvature quadrupole (BCQ) can be integrated by parts twice to the form 
\begin{equation}\label{eq:qint}
        Q_{abc} = \int_{\bm{k}} \Omega_{\bm{k},c} (\partial^2_{ab} f_0). 
\end{equation}
It is important to note that for a lattice model, Eq.~(\ref{eq:qorig}) and Eq.~(\ref{eq:qint}) are always equivalent, as already discussed in the context of BCD~\cite{PhysRevLett.123.246602}. However, in the case of a low-energy model, one should add a correction to Eq.~(\ref{eq:qint}) which we find to be vanishing in the system under consideration. 
Given the above equation, it is clear that $\lambda^{\kappa,1}$ in Eq.~(\ref{eq:lambda_kappa}) is generated due to a modulated BCQ. 

In the context of symmetries, the modulated BCQ is odd under time-reversal symmetry, leading to the fact that the anomalous thermal Hall effect mediated by $\lambda^{\kappa,n}$ vanishes in time-reversal invariant systems. This becomes apparent when we examine Eq.~(\ref{eq:lambda_kappa}) under time-reversal symmetry $\mathcal{T}$: $\partial_a \xrightarrow{} -\partial_a $, $\epsilon_{\bm{k}}=\epsilon_{\bm{-k}}$ and $\Omega_{\bm{k}} \xrightarrow{} -\Omega_{\bm{k}}$, resulting in $\lambda_{abc} \xrightarrow{} -\lambda_{abc}$. In addition, discrete crystalline symmetries such as rotation and reflection play an important role in how the BCQ and modulated BCQ transform. For example, in 2D space, a rotation by an angle $\theta_P$ about the $z$-axis is captured by the unitary operator $\hat{R}_z(\theta_P)=exp[-i \, \frac{\hat{L}_z}{\hbar} \, \theta_P]$ where $\hat{L}_z$ is the generator of rotations, namely, the angular momentum operator~\cite{Manna_2019}. Eigenvectors $Q_m$ of $\hat{L}_z$ can be constructed from linear combinations of the three independent components $Q_{abz}$ of the BCQ. Under $p$-fold rotation, we have $\hat{R}_z(2\pi/p)Q_m=e^{-i2\pi m/p}Q_m$, where $m=0, \pm 2$ is the eigenvalue of $\hat{L}_z$ divided by $\hbar$. Here, $Q_{0}$, and $Q_{\pm 2}$ represent the trace and traceless component of BCQ, respectively. Since the BCQ has to be invariant under this rotation, we have $(e^{-i2\pi m/p}-1)Q_m=0$. This leads to the fact that any rotational symmetry with an order $p$ higher than the spatial dimension ($m = 2$ here) of the system forces the BCQ to vanish since $e^{-i2\pi m/p} \neq 1$ for $p > |m|$. Despite the different energy modulations, it is clear that all the $\lambda^{\kappa,n}_{cde}$'s transform exactly like $Q_{abc}$ under symmetry operations~\cite{Zhang_2023} and we henceforth refer to them as the thermal counterparts of Berry curvature quadrupole (TCBCQ). This leads to the thermal Hall and Nernst (Sec. \ref{ANE}) contributions mediated by the TCBCQ similarly vanishing in time-reversal invariant systems. 

To find a candidate two-dimensional (2D) system to observe the leading-order TCBCQ-induced third-order thermal Hall and Nernst effects based on symmetry analysis, we point out that there are 31 magnetic point groups (MPGs) in 2D space violating time-reversal symmetry \cite{Zhang_2023, Newnham1988}. However, along with the BCQ, the BCM (i.e., $\int_{\bm{k}}\Omega_{\bm{k}}f_0$) as well as the BCD (i.e., $D_a=\int_{\bm{k}}(\partial_a \Omega_{\bm{k}})f_0$) can simultaneously appear in $\mathcal{T}$-broken systems. Specifically, in a 2D system, the largest symmetry that allows for a finite BCD is the presence of a single mirror symmetry~\cite{Sodemann_2015}. Now, in 10 of these MPGs, the Berry curvature vanishes identically throughout the Brillouin zone due to the presence of $\mathcal{C}_2\mathcal{T}$ symmetry~\cite{Bradlyn_2019}, leading to vanishing Berry curvature multipole-mediated anomalous Hall responses. Among the remaining 21 MPGs, 10 MPGs host a nonvanishing BCM because of broken time-reversal and mirror symmetries. Now, 3 of the remaining 11 MPGs contain a single mirror line, therefore allowing a nonvanishing BCD. Since BCQ is finite in 2D systems that respect rotational symmetry with order lower than 2, only 3 MPGs among the remaining 8 (e.g. $2mm$) can have BCQ/TCBCQ-induced leading-order responses~\cite{Zhang_2023}.  

\subsection{\label{ANE}Anomalous Nernst Effect}
The anomalous Nernst effect refers to the generation of a transverse charge current in the presence of a longitudinal temperature gradient due to a Berry-phase-induced intrinsic mechanism. While the anomalous charge Hall effect is mediated by the anomalous velocity produced by a mechanical force (arising from $\bm{E}$), the anomalous Nernst response is mediated by a Berry-phase correction term to the orbital magnetization in the presence of a statistical force (arising from $-\bm{\nabla} T$) \cite{Xiao_2010, Niu_2006}. The general expression of transverse Nernst-like charge current in the presence of a longitudinal thermal gradient for a system with non-zero Berry curvature is given by \cite{Niu_2006} 
\begin{equation}
\begin{split}
     \bm{j^E_A} =& -\frac{e}{\hbar} \bm{\nabla} T \times \int_{\bm{k}}  \bm{\Omega}^m_{\bm{k}} \biggl[ \frac{(\epsilon^m_{\bm{k}}-\mu)}{T} f^m_0 \\
        &+ k_BT \ln(1+e^{-\beta(\epsilon^m_{\bm{k}}-\mu)}) \biggr].   
\end{split}
\end{equation}
Plugging in the expression for $f_2$ in the above equation as we did in deriving Eq.~(\ref{eqj3}), the third-order Nernst-like current density can be written as 
\begin{equation}
\begin{split}
     (j^E_A)_{a,3} = &-\alpha^{23}_{abde} (\nabla_b T)(\nabla^2_{de}T)\\
     &- \alpha^{222}_{abde} (\nabla_b T)(\nabla_d T)(\nabla_e T).\\
 \end{split}
 \end{equation}
 Here, the superscript indices of the coefficients follow the same notion as for Eq.~(\ref{eqj3}). The third-order Nernst coefficients are expressed as
 \begin{equation}\label{eq:alpha23}
     \alpha^{23}_{abde} = \varepsilon_{abc}\tau^2\frac{e k_B^3T}{\hbar^3}\lambda^{\alpha,2}_{cde},\\
 \end{equation}
 and
\begin{equation}\label{eq:alpha222}
    \alpha^{222}_{abde} = -\varepsilon_{abc}\tau^2\frac{ek_B^3}{\hbar^3}\biggl[ \lambda^{\alpha,1}_{cde} + 2\lambda^{\alpha,2}_{cde} - \lambda^{\alpha,3}_{cde} \biggr],\\
\end{equation}
where
\begin{equation}\label{eq:lambda_alpha}
\begin{split}
    \lambda^{\alpha,1}_{cde} = &\int_{\bm{k}}\Omega_{\bm{k},c}\frac{(\epsilon_{\bm{k}}-\mu)^3}{(k_{B}T)^3}  \bigl(\partial^2_{de} f_0)\\    
    \lambda^{\alpha,2}_{cde} = &\int_{\bm{k}}\Omega_{\bm{k},c}\frac{(\epsilon_{\bm{k}}-\mu)^2}{(k_{B}T)^3} (\partial_d \epsilon_{\bm{k}})(\partial_e f_0)\\
    \lambda^{\alpha,3}_{cde} = &\int_{\bm{k}}\Omega_{\bm{k},c}\frac{(\epsilon_{\bm{k}}-\mu)^3}{(k_{B}T)^3} \bigl(\partial^2_{de} \epsilon_{\bm{k}}) (\frac{\partial f_0}{\partial \epsilon_{\bm{k}}}\bigr).
\end{split}
\end{equation}
 Similar to the thermal Hall conductivities, the Nernst conductivities $\alpha^{23}$ and $\alpha^{222}$ are Fermi surface quantities due to the presence of $\partial f_0/\partial\epsilon_{\bm{k}}$, and they, too, scale quadratically with scattering time $\tau$. Similar to $\lambda^{\kappa,n}_{cde}$'s, $\lambda^{\alpha,n}_{cde}$'s represent the Berry curvature quadrupole with different energy modulations and follow the same symmetry operations mentioned above.
 To find the low-temperature dependence of  $\alpha^{23}$ and $\alpha^{222}$ in the low-temperature limit, we apply Sommerfeld expansion to $\lambda^{\alpha,1}$, $\lambda^{\alpha,2}$ and $\lambda^{\alpha,3}$ and arrive at
\begin{equation}\label{eq:T_a23}
     \alpha^{23}_{abde}(T,\mu) = \varepsilon_{abc}\frac{\pi^2}{6}\frac{ek_B^2}{\hbar^3}\tau^2 a^{(2)}_{cde}(\mu) + O(T^2),
 \end{equation}
 where 
\begin{equation}
    a_{cde}(\epsilon) = -\int_{\bm{k}} \Omega_{\bm{k},c}\partial_{d}\epsilon_{\bm{k}}\partial_{e}\epsilon_{\bm{k}}(\epsilon-\mu)^2\delta(\epsilon-\epsilon_{k}), \nonumber
\end{equation}
and 
\begin{equation}\label{eq:T_a222}
    \alpha^{222}_{abde}(T,\mu) = -\varepsilon_{abc}\frac{\pi^2}{6}\frac{ek_B^2}{\hbar^3}\frac{1}{T}\tau^2  A^{(2)}_{cde}(\mu) + O(T),
\end{equation}
where 
\begin{equation}
    A_{cde}(\epsilon) = \int_{\bm{k}} \Omega_{\bm{k},c}\partial_{d}\epsilon_{\bm{k}}\partial_{e}\epsilon_{\bm{k}}\biggl[(\epsilon-\mu)^4\frac{\partial\delta}{\partial \epsilon} + 2(\epsilon-\mu)^3\delta(\epsilon-\epsilon_{k})\biggr]. \nonumber
\end{equation}
Here, $a^{(n)}_{cde}(\mu) = \partial^{n}a_{cde}/\partial \mu^n$, and $A^{(n)}_{cde}(\mu) = \partial^{n}A_{cde}/\partial \mu^n$. Therefore, it is clear that $\alpha^{23} \propto T^0$ and $\alpha^{23} \propto T^{-1}$ in the low-temperature limit.
 The schematic diagram of the experimental setup for measuring third-order anomalous thermal Hall and Nernst effects is shown in Fig~\ref{fig:exp}.

\begin{figure}[tb]
\centering
\includegraphics[width=0.485\textwidth]{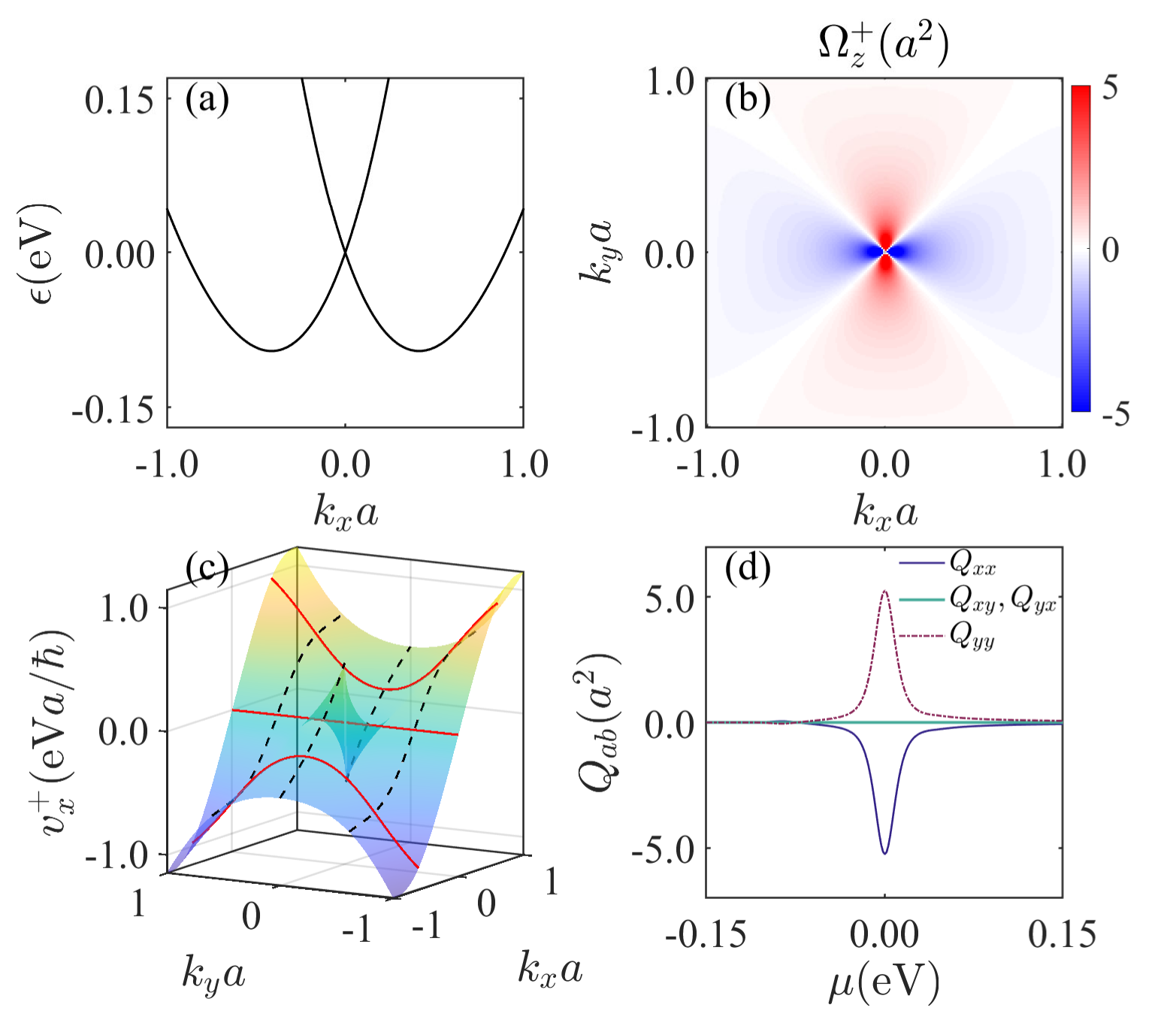}
\caption{ The energy dispersion along the $k_x$-direction ($k_y = 0$) of the effective model given in Eq.~(\ref{eq:model}) is depicted in (a). (b) shows the Berry curvature distribution ($\Omega_{z}^{+}$) of the conduction band. (c) depicts the carrier velocity profile along the $x$-direction for the conduction band ($v^+_x$) as given by the semiclassical equations of motion showing that under $\mathcal{C}_{4}\mathcal{T}$ symmetry, the velocity is odd in momentum space, $v^+_x (\bm{k}) = -v^+_x (-\bm{k})$. The solid red lines show that the profile is an even function w.r.t. $k_y$, and the dashed black lines show it is odd w.r.t. $k_x$. (d) The doping dependence $\mu$ (can be tuned by gate voltage) of the different components of the Berry curvature quadrupole are shown.  Here, we consider $t=3$, $v=1$, $m=2$, and length scale $a = m/v$.}
\label{fig:hamiltonian}
\end{figure}

\begin{figure*}[h!tb]
\centering
\includegraphics[width=0.98\textwidth]{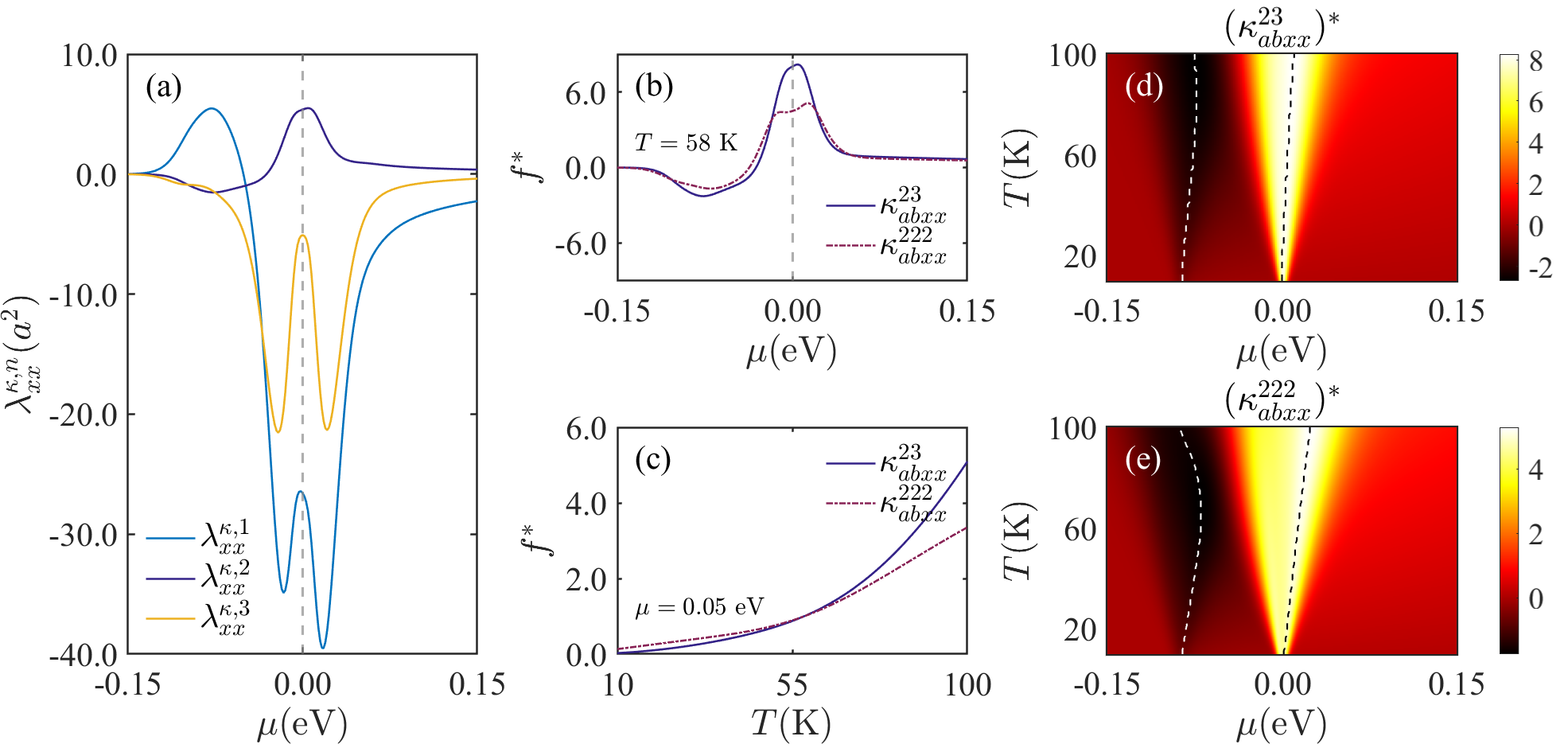}
\caption{For effective model given in Eq.~(\ref{eq:model}), the thermal counterparts of Berry curvature quadrupole $\lambda^{\kappa,n}_{cde}$'s associated with $(j^Q_A)_{a,3}$ as a function of doping $\mu$ for a particular temperature $k_{B}T = 0.005$ eV are depicted in (a). Recall that for this 2D system, there is only one independent component $\lambda^{\kappa,n}_{zxx} = -\lambda^{\kappa,n}_{zyy}$ and we have dropped the $c\equiv z$ index for brevity. The variation of the third-order anomalous thermal Hall coefficients $\kappa^{23}_{abxx}$ and $\kappa^{222}_{abxx}$ is depicted as a function of (b) $\mu$ ranging from $-0.15$ eV to $0.15$ eV, (c) $T$ ranging from 10-100K. (d)-(e) shows the colormap of $\kappa^{23}_{abxx}$ and $\kappa^{222}_{abxx}$ as  a function of both $\mu$ and $T$, respectively. The white line is the locus of the minima, and the black line is the locus of the maxima. The chemical potential of the maxima increases with increasing temperature. $f^*$ indicates that the function $f$ has been scaled with respect to its value at $\mu=0.05$ eV and $k_BT = 0.005$ eV ($T\sim58$K).}
\label{fig:ThermalHall}
\end{figure*}

\section{\label{results}Results}
To investigate the third-order anomalous thermal Hall and Nernst responses, we consider an effective Hamiltonian of a two-dimensional system belonging to the magnetic point group $4'm'm$, which up to the second-order in $k$ near the $\Gamma$ point can be written as~\cite{Zhang_2023} 
\begin{equation}\label{eq:model}
    \mathcal{H}(\bm{k}) = tk^{2} + v(k_{y}\sigma_{x}-k_{x}\sigma_{y}) + m(k_{x}^{2}-k_{y}^{2})\sigma_{z},
\end{equation}
where $\sigma_x, \sigma_y, \sigma_z$ denote the Pauli matrices acting on the spin degrees of freedom, and $k = |\bm{k}|$. Here, the first term $tk^{2}$ represents the kinetic energy, and the third term generates a second-order warping and breaks time-reversal symmetry in spin space. The second term denotes a Rashba interaction and couples the spin and orbital spaces. The two generators of the $4'm'm$ group are $\mathcal{C}_{4}\mathcal{T}$ and $\mathcal{M}_{x}\mathcal{T}$, where $\mathcal{C}_{4}$ is the four-fold rotation around the z-axis, $\mathcal{T}$ denotes the time-reversal symmetry, and $\mathcal{M}_{x}$ is a reflection with respect to the $yz$-plane $x \xrightarrow{} -x$. The band touching at $\Gamma$-point is protected by the $\mathcal{C}_{4}\mathcal{T}$ symmetry,  the combination of $\mathcal{C}_{4}$ and time-reversal symmetries. We would like to point out that the monolayer SrMnBi$_2$, which is an antiferromagnet, belongs to the $4'm'm$ group. Diagonalizing the above Hamiltonian, the energy dispersion can be obtained as
\begin{equation}\label{eqen}
    \epsilon_{\bm{k}}^\pm = tk^{2} \pm |\bm{d(k)}|,
\end{equation}
where $\pm$ denote the conduction and valence bands, respectively, and $\bm{d(k)}=[vk_{y},-vk_{x},m(k_{x}^{2}-k_{y}^{2})]$. The band dispersion is shown in Fig.~\ref{fig:hamiltonian}(a).

We would now like to discuss this system's symmetry-enforced leading-order anomalous thermal Hall and Nernst effects. Since the considered system is in 2D space, the Berry curvature $\bm{\Omega}$ is reduced from a pseudovector to a pseudoscalar with only the non-zero component being $\Omega_{(c=z)}$, i.e., the component perpendicular to the $xy$-plane. 
Using the energy eigenket and eigenvalues, the Berry curvature for the 2D system under consideration is evaluated as~\cite{Xiao_2010}
\begin{equation}
    \Omega^{n}_{\bm{k},z} = i\sum_{n'\neq n} \frac{\bra{n}dH/dk_{\alpha}\ket{n'}\bra{n'}dH/dk_{\beta}\ket{n}-(\alpha\leftrightarrow\beta)}{(\epsilon^{n}_{\bm{k}}-\epsilon^{n'}_{\bm{k}})^{2}}
\end{equation}
where $n$ is the band index, $\ket{n}$ is the energy eigenket and $\alpha,\beta$ represent components of the vector. 
The $z$-component of the Berry curvature is obtained as
\begin{equation}\label{eqbc}
    \Omega^{\pm}_{\bm{k},z} = \mp \frac{v^{2}d_{z}(\bm{k})}{2|\bm{d(k)}|^{3}}.
\end{equation}
\begin{figure*}[t]
\centering
\includegraphics[width=0.98\textwidth]{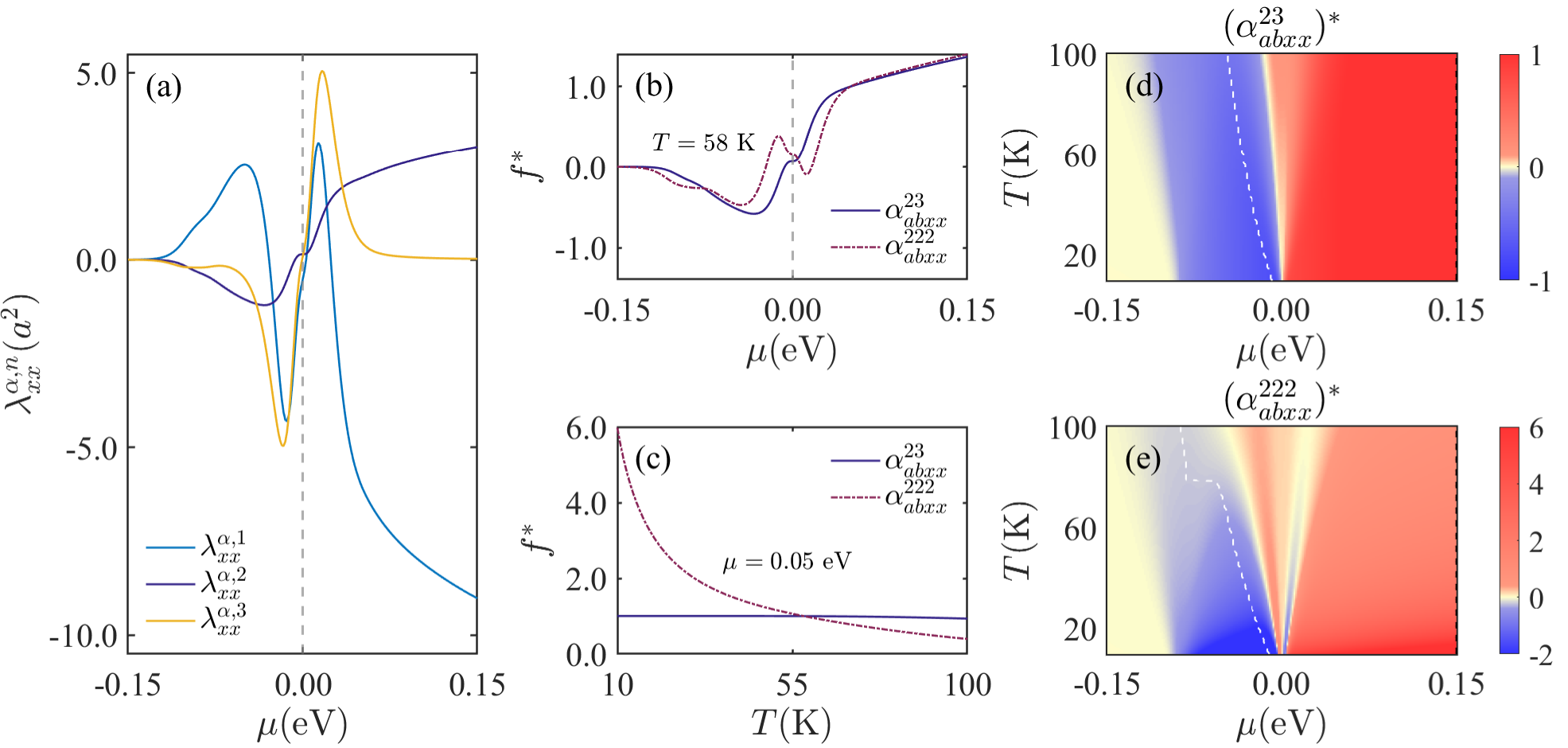}
\caption{For effective model given in Eq.~(\ref{eq:model}), the thermal counterparts of Berry curvature quadrupole $\lambda^{\alpha,n}_{cde}$'s associated with $(j^E_A)_{a,3}$ as a function of doping $\mu$ for $k_BT = 0.005$ eV are depicted in (a). The variation of the third-order anomalous Nernst coefficients $\alpha^{23}_{abxx}$ and $\alpha^{222}_{abxx}$ is depicted as a function of (b) $\mu$ ranging from $-0.15$ eV to $0.15$ eV, (c) $T$ ranging from 10-100K. (d)-(e) shows the colormap of $\alpha^{23}_{abxx}$ and $\alpha^{222}_{abxx}$ as  a function of both $\mu$ and $T$, respectively. The white line is the locus of the minima, and the black line is the locus of the maxima. The chemical potential of the maxima increases with increasing temperature. $f^*$ indicates that the function $f$ has been scaled with respect to its value at $\mu=0.05$ eV and $k_BT = 0.005$ eV ($T\sim58$K).}
\label{fig:Nernst}
\end{figure*}
It is important to note that Berry curvature multipole-induced first-order, second-order and third-order thermoelectric responses can appear simultaneously in experiment for a system with broken time-reversal and inversion symmetries. However, for the system considered here, the linear anomalous electric/thermal Hall effect mediated by the Berry curvature monopole vanishes due to the presence of $\mathcal{C}_{4}\mathcal{T}$ symmetry. In addition, the Berry curvature dipole also vanishes due to $\mathcal{C}_2 = (\mathcal{C}_{4}\mathcal{T})^2$. The Berry curvature profile of the system Hamiltonian is depicted in Fig.~\ref{fig:hamiltonian}(b). It is clear from Fig.~\ref{fig:hamiltonian}(a)-(b) that under $\mathcal{C}_{4}\mathcal{T}$ symmetry, the Berry curvature and the energy dispersion satisfy the relations $\Omega_{\bm{k},z} = \Omega_{-\bm{k},z}$, $\epsilon_{\bm{k}} =\epsilon_{-\bm{k}}$. In addition, the group velocity $v_d(k_x,k_y) =  \hbar^{-1} \partial_d \epsilon_{\bm{k}} $ (where $d \in {x,y}$ ), shown in Fig.~\ref{fig:hamiltonian}(c), is an odd function with respect to $k_d$ and an even function with respect to $k_{d\perp}$, making it odd in $\bm{k}$. This makes $\partial_d f_0 = \hbar (\partial f_0/ \partial \epsilon_{\bm{k}}) v_d$ an odd function in $\bm{k}$, and $\partial^2_{de} f_0$ even in $\bm{k}$ when $d=e$ and odd when $d \neq e$. Examining Eq.~(\ref{eq:qint}), it is now clear that the integrand is odd in $\bm{k}$ for $d \neq e$, making $Q_{zxy} = 0 = Q_{zyx}$, leaving one independent component of Berry curvature quadruple tensor  $Q_{xx} = -Q_{yy}$ (index $z$ is suppressed in a 2D system). The profile of the Berry curvature quadrupole as a function of doping is shown in Fig.~\ref{fig:hamiltonian}(d). Based on the above analyses, we conclude that the symmetries of the system ensure that the first- and second-order anomalous thermoelectric and thermal Hall responses vanish. Therefore, the modulated Berry curvature quadrupole-mediated third-order anomalous thermal and thermoelectric responses will be the leading order response in this system.

Having calculated the BCQ, we now investigate the thermal counterparts of BCQ and the third-order anomalous thermoelectric and thermal Hall coefficients they induce. Since the thermal counterparts of BCQ, $\lambda_{zde}$'s follow the same symmetry requirements as $Q_{zde}$, it is clear that despite varying degrees of energy modulation, only one independent component of $\lambda_{zde}$ will survive in this system. In particular, $\lambda_{xx} = -\lambda_{yy}$ with $\lambda_{xy} = 0 = \lambda_{yx}$. This naturally leads to one independent component for the third-order thermal Hall and Nernst coefficients of this system, as seen from Eqs.~(\ref{eq:lambda_kappa}) and (\ref{eq:lambda_alpha}):   
$\kappa_{abxx} = -\kappa_{abyy}$ and $\alpha_{abxx} = -\alpha_{abyy}$, with $\kappa_{abxy} = 0 =\kappa_{abyx}$ and $\alpha_{abxy} = 0 =\alpha_{abyx}$.  

The numerically evaluated TCBCQs $\lambda^{\kappa,n}$'s with different energy modulations are depicted in Fig.~\ref{fig:ThermalHall}(a) as a function of chemical potential $\mu$. It is important to note that the energy dispersion of the system near $\mu=0$ (band-touching point) can be approximated as $E^{\pm}(\bm{k}) \approx \pm v \bm{k}$. This leads to the Berry curvature being of the form $\Omega_{z}^{\pm} = \mp d_z/v|\bm{k}|^3$. It follows that the Berry curvatures of both bands exhibit a large value near the band-crossing point, as shown in Fig.~\ref{fig:hamiltonian}(b), which is inherited by the BCQ. However, the behavior of the TCBCQs (seen in Figs.~\ref{fig:ThermalHall}(a) and \ref{fig:Nernst}(a)) is more complicated owing to the interplay between the varying degrees of energy modulation and the orders of energy derivatives involved. We note that in contrast to the Berry curvature, the $\lambda^{\kappa,n}$'s carry the same sign for conduction and valence band for a two-band model system. It is clear from the figure that the TCBCQ $\lambda^{\kappa,n}$'s exhibit peaks near $\mu=0$, i.e., the band-crossing point, irrespective of the degree of modulation. When we tune the chemical potential away from the band-touching point, the magnitude of $\lambda^{\kappa,2}$ decreases gradually while the magnitudes of $\lambda^{\kappa,1}$ and $\lambda^{\kappa,3}$ briefly increase before decreasing as we move farther away from the band-touching point. Interestingly, we find $\lambda^{\kappa,2}$ has the opposite sign compared to $\lambda^{\kappa,1}$ and $\lambda^{\kappa,3}$ for $\mu \geq 0$.
This can be understood by considering the integrands in Eq.~(\ref{eq:lambda_kappa}). Near the band-crossing point, for both bands, the integrand of $\lambda^{\kappa,2}$ bears $\Omega_{\bm{k},z}(\epsilon_{\bm{k}}-\mu)(\partial_x \epsilon_{\bm{k}})(\partial_x f_0) \equiv  \Omega_{\bm{k},z}(\mu - \epsilon_{\bm{k}})v_x^2 \delta(\epsilon_{\bm{k}}-\mu)$ which carries a net positive sign, whereas $\Omega_{\bm{k},z}(\partial^2_{xx} f_0)$ and $\Omega_{\bm{k},z}\bigl(\partial^2_{xx} \epsilon_{\bm{k}}) (\frac{\partial f_0}{\partial \epsilon_{\bm{k}}}\bigr)$ in the integrands of $\lambda^{\kappa,1}$ and $\lambda^{\kappa,3}$, respectively carry net negative signs.
In addition to the cusp (peak) near $\mu = 0$, $\lambda^{\kappa,1}$ ($\lambda^{\kappa,2}$) shows a peak (cusp) around the band-minimum $\mu\sim -0.096$ eV. 
We note that the energy modulation increases the magnitude of the TCBCQs, as seen in Fig.~\ref{fig:ThermalHall}(a).

The calculated $abxx$-component of the conductivities $\kappa^{23}$ and $\kappa^{222}$ of third-order anomalous thermal Hall response as a function of the chemical potential $\mu$ for $k_{B}T=0.005$ eV are shown in Fig.~\ref{fig:ThermalHall}(b). It is clear from the figure that $\kappa^{23}$ follows the same behavior of $\lambda^{\kappa, 2}$ as evident from Eq.~(\ref{eq:k23}). On the other hand, $\kappa^{222}$ contains three different Berry curvature quadrupole contributions. However, it is dominated by the $\lambda^{\kappa, 1}$ as the magnitudes of $\lambda^{\kappa, 2}$ and $\lambda^{\kappa, 3}$ are smaller compared to $\lambda^{\kappa, 1}$, as shown in Fig.~\ref{fig:ThermalHall}(a). In Fig.~\ref{fig:ThermalHall}(c), we have shown the temperature dependence of both $\kappa^{222}$ and $\kappa^{23}$ for a fixed $\mu=0.05$ eV. We find that both the coefficients increase with increasing temperature. Specifically, at low temperatures (below $55$ K), $\kappa^{23}$ shows a $\sim T^2$ dependence and $\kappa^{222}$ shows a $\sim T$ dependence as expected from Eq.~(\ref{eq:T_k23}) and Eq.~(\ref{eq:T_k222}), respectively.  
Finally, in Fig.~\ref{fig:ThermalHall}(d) and Fig.~\ref{fig:ThermalHall}(e), we have shown $\kappa^{23}$ and $\kappa^{222}$ as a function of both $\mu$ and $T$ respectively. The figures show that as we tune $\mu$ from the band-touching point towards the top of the conduction band, the magnitude of both coefficients decreases at all temperatures. On the other hand, when we tune the chemical potential from $\mu=0$ towards the bottom of the valence band, we see a cusp similar to the one seen in Fig.~\ref{fig:ThermalHall}(b) for all temperatures. As the temperature increases, the width of the Fermi-Dirac distribution increases and more energy levels contribute to the integrand, leading to all the peaks increasing in width as well. 

We now turn our attention to the quantities pertaining to the third-order anomalous Nernst effect. Fig.~\ref{fig:Nernst} shows the model-specific calculated results of the general equations in Eqs.~(\ref{eq:alpha23})--(\ref{eq:lambda_alpha}) for the Hamiltonian given by Eq.~(\ref{eq:model}). Immediately, we notice that the $\alpha$'s in Fig.\ref{fig:Nernst}(b) increase with $\mu$ whereas the $\kappa$'s in Fig.~\ref{fig:ThermalHall}(b) decreased with $\mu$. Comparing Eqs.~(\ref{eq:lambda_alpha}) with~(\ref{eq:lambda_kappa}), we find that the distinction between the two sets of integrands is that the ones for $\lambda^{\alpha,n}$ are one power lower in $(\epsilon_{\bm{k}} - \mu)/k_B T$ than their corresponding $\lambda^{\kappa,n}$ counterparts. Consequently, we expect the TCBCQs associated with the third-order anomalous Nernst current to show not only a difference in magnitude compared to their thermal Hall counterparts but also a change in sign as we traverse the band crossing point at $\mu = 0$ due to an effective odd power of $\mu$. Naively, one would expect the sign change to occur when the Fermi energy is in the valance band ($\mu < 0$), but Fig.~\ref{fig:Nernst}(a) shows the sign changing when the Fermi energy is in the conduction band  ($\mu > 0$). Combined with the band structure shown in Fig.~\ref{fig:hamiltonian}(a), $\Omega_{\bm{k}}$ for both the bands being sharply peaked at the band crossing point results in the integrands having a greater contribution from levels above the Fermi energy $(\epsilon_{\bm{k}} - \mu)>0$ in the valance band, and from levels below the Fermi energy $(\epsilon_{\bm{k}} - \mu)<0$ in the conduction band. We note that similar to the Berry curvature and in contrast to $\lambda^{\kappa,n}$'s, the $\lambda^{\alpha,n}$'s carry opposite signs for conduction and valence bands for a two-band system. 

In Fig.~\ref{fig:Nernst}(b), we see the  behavior of $\lambda^{\alpha,n}$'s being extended to $\alpha^{23}_{abxx}$ and $\alpha^{222}_{abxx}$, with the conductivites changing sign for $\mu>0$. The temperature dependence of $\alpha$'s depicted in Fig.~\ref{fig:Nernst}(c) shows them to be decreasing in magnitude with increasing $T$, once again contrasting with the behavior of $\kappa$'s, which are seen to be increasing with $T$ in Fig.~\ref{fig:ThermalHall}(c). Thermal broadening of the Fermi distribution explains the decrease in magnitude when the Fermi level is in the conduction band ($\mu = 0.05$ eV) and close to the band-touching point; as the distribution picks up more and more energy levels, the integrand collects more contributions from the levels below the Fermi energy $(\epsilon_{\bm{k}} - \mu) < 0$ (as the levels are closer to the band crossing point and thereby have a larger Berry curvature) than from the levels above $(\epsilon_{\bm{k}} - \mu) > 0$ which results in a net negative contribution. The temperature dependence for the $\alpha$'s comes from $1/T^3$ as well as the broadening of $\partial f_0/\partial \epsilon_{\bm{k}}$. The result is that for low temperatures ($T<55$ K), $\alpha^{23}$ has a very weak temperature dependence $\sim T^0$ as expected from Eq.~(\ref{eq:T_a23}) while $\alpha^{222}$ shows a $\sim 1/T$ dependence according to Eq.~(\ref{eq:T_a222}). Finally, in Fig.~\ref{fig:Nernst}(d) and Fig.~\ref{fig:Nernst}(e), we have shown $\alpha^{23}$ and $\alpha^{222}$ as a function of both $\mu$ and $T$, respectively. 

We would like to point out that identifying the non-zero tensor components of TCBCQ for other possible Hamiltonians is pretty straightforward, following the symmetry analysis in Sec.~\ref{ATHE}. The three independent components of TCBCQ, which follow the same constraints as the corresponding BCQ, are $\lambda_{xx}$, $\lambda_{xy}$ and $\lambda_{yy}$. The eigenvectors of $\hat{L}_z$ constructed from linear combinations of these components are $\lambda_{+2}=\int_{\bm{k}}f_{0}\partial^{2}_{+}\Omega$, $\lambda_{0}=\int_{\bm{k}}f_{0}\partial_{+}\partial_{-}\Omega$, $\lambda_{-2}=\int_{\bm{k}}f_{0}\partial^{2}_{-}\Omega$ where $\partial_{\pm}=\partial_{x}\pm i\partial_{y}$. Specifically, 
\begin{equation}\label{eq:lpm}
    \begin{split}
        \lambda_{+2} = & \lambda_{xx} - \lambda_{yy} + 2i\lambda_{xy}\\
        \lambda_{0} = & \lambda_{xx} + \lambda_{yy}\\
        \lambda_{-2} = & \lambda_{xx} - \lambda_{yy} - 2i\lambda_{xy}.\\
    \end{split}
\end{equation}
We now analyze the components of TCBCQ in quantum anomalous Hall state by considering the general Hamiltonian \cite{PhysRevB.74.085308, PhysRevLett.101.146802}
\begin{equation}
   \mathcal{H}(\bm{k}) = tk^2 + v(k_{y}\sigma_{x}-k_{x}\sigma_{y}) + M(\bm{k})\sigma_{z}, 
\end{equation}
where $tk^2$ is the kinetic energy term and $M(\bm{k}) = M_{0}-M_2k^2$ is the mass term giving the quantum anomalous Hall phase when $M_0 M_2 > 0$. The model has a continuous rotational symmetry, meaning that for a rotation through any angle $\theta$, we have $\hat{R}_z(\theta)\lambda_m=e^{-i\theta m}\lambda_m$, requiring $e^{-i\theta m}\lambda_m = \lambda_m$. Here, $\lambda_m$'s (with $m = 0, \pm2$) are the eigenvectors of $\hat{L}_z$ constructed from linear combinations of the three independent components $\lambda_{abz}$ of the TCBCQ. Since for arbitrary $\theta$ we do not have $e^{-i\theta m} = 1$, we have necessarily $\lambda_{\pm 2} = 0$. Therefore, Eq.~(\ref{eq:lpm}) immediately gives $\lambda_{xx} = \lambda_{yy}$ and $\lambda_{xy} = 0$. 
 
\section{\label{conclusions} Conclusions}
In this work, within the semiclassical Boltzmann0 transport formalism framework with momentum-independent relaxation-time approximation, we analytically derive the general expression of third-order anomalous thermal Hall and Nernst current density. We show that they are mediated by the energy-modulated TCBCQs, which require the breaking of time-reversal symmetry to be finite. We find that both the thermal Hall and the Nernst coefficients scale quadratically with relaxation time $\tau$ (see Eqs.~(\ref{eq:k23})--(\ref{eq:k222}) for thermal Hall and Eqs.~(\ref{eq:alpha23})--(\ref{eq:alpha222}) for Nernst). Using symmetry analysis, we find that the TCBCQs follow the same symmetry requirements as the unmodulated BCQ and could consequently appear as a leading order Berry curvature moment in three two-dimensional magnetic point groups: $2mm$, $4'$, $4'm'm$. Based on the above discussion, we investigate the third-order anomalous thermal Hall and Nernst responses in a two-dimensional Rashba-like system (possible candidate material: monolayer SrMnBi$_2$) with second-order Fermi surface warping belonging to the $4'm'm$ magnetic point group. In this system, the Berry curvature monopole mediated linear anomalous thermal Hall and Nernst effects vanish due to the presence of $\mathcal{C}_{4}\mathcal{T}$ symmetry. In addition, the Berry curvature dipole vanishes due to $\mathcal{C}_2 = (\mathcal{C}_{4}\mathcal{T})^2$, as do the corresponding second-order thermal Hall and Nernst effects. 

We make experimental predictions for the TCBCQ-mediated anomalous third-order thermoelectric coefficients and their dependence on temperature and chemical potential for a 2D Rashba-like system with second-order Fermi surface warping. At a constant temperature, we show that third-order thermal Hall coefficients exhibit peaks near the band-touching point and decrease gradually as the chemical potential is tuned away from the band-touching point. On the other hand, third-order Nernst coefficients encounter a change in sign as we cross the band-touching point. We have also extracted the temperature scaling at low temperatures for both thermal Hall and Nernst effects. To obtain a quantitative order of magnitude estimate of the third-order thermal Hall and Nernst responses, we assume a uniform external thermal gradient applied along the $x$-direction. Considering $\nabla_x T = 1$ K/mm \cite{Guin2019-xq}, $\tau = 10^{-11} s$, $\mu = 0.05$ eV, $T = 58$ K, the thermal Hall current in the $y$ direction is given by $ (j^Q_A)_{y,3} = -\kappa^{222}_{yxxx} (\nabla_x T)^3 = 8.68 \times 10^{-23}$ W/mm, and the Nernst current in the y direction by $ (j^E_A)_{y,3} = -\alpha^{222}_{yxxx} (\nabla_x T)^3 = 1.22 \times 10^{-20}$ A/mm. We discount the contributions to the currents from $\kappa^{23}_{abde} \nabla_b T \nabla^2_{de} T$ and $\alpha^{23}_{abde} \nabla_b T \nabla^2_{de} T$ by assuming that in experiment, the temperature gradient can be made to have a sufficiently small spatial dependence ($\nabla^2_{de} T \sim 0$). Finally, with their characteristic behaviors, these higher-order nonlinear anomalous thermal Hall and Nernst responses could serve as promising tools for probing band geometric quantities in novel materials with low crystalline symmetry in experiments. We note that although we expect the third-order thermal Hall and Nernst responses to dominate in some specific magnetic systems based on the above symmetry analysis, the linear and second-order signals can also appear simultaneously in several other systems. The responses of different orders can be distinguished by their temperature and chemical potential dependencies in experiment. 

Along with the TCBCQ-induced third-order anomalous thermal Hall and Nernst current, disorder-mediated contributions (nonlinear side jump and skew-scattering contributions)  and Berry-curvature-independent contributions may exist. Although the Berry-curvature-independent contribution vanishes for third-order transverse thermoelectric coefficients, the disorder-mediated contributions could be essential to explain the recent ongoing experiments as was the case for second-order anomalous responses~\cite{Nandy_2019_ES}. Therefore, investigating the third-order disorder-mediated (side-jump and skew-scattering) contributions to the anomalous thermal Hall and Nernst effects could be an exciting direction for future study. In addition, the Wiedemann-Franz law and Mott relation for the BCD-induced second-order anomalous response exhibit different forms compared to the linear response case~\cite{Zeng_2020}. In view of this, deriving these fundamental relations connecting anomalous third-order charge Hall, Nernst and thermal Hall effects is a fascinating direction to study.
\section{Acknowledgements} S.K. and S.T. acknowledge support from ARO Grant No. W911NF2210247 and ONR Grant No. N00014-23-1-2061. The work at Los Alamos National Laboratory was carried out under the auspices of the U.S. Department of Energy (DOE) National Nuclear Security Administration under Contract No. 89233218CNA000001. It was supported by the LANL LDRD Program and in part by the Center for Integrated Nanotechnologies, a DOE BES user facility, in partnership with the LANL Institutional Computing Program for computational resources.
\bibliography{TOTHE} 
\end{document}